\documentclass[final,5p,times,twocolumn]{elsarticle}

\usepackage{amssymb}
\usepackage{amsmath}

\usepackage{dcolumn}%
\usepackage{bm}%
\usepackage{graphicx,textcomp}
\usepackage{array}
\usepackage{bm}
\usepackage{mathtools}
\usepackage{amssymb,amsfonts,amsmath}
\usepackage{color}
\usepackage{nicefrac}
\usepackage{algorithm}
\usepackage{algpseudocode}
\usepackage{enumerate}
\usepackage[all]{nowidow}
\usepackage{booktabs}
\usepackage{multirow}
\usepackage{xcolor}

\usepackage{lineno}
\journal{Chaos, Solitons \& Fractals}

\begin{document}
	
	\begin{frontmatter}
		
		
		
		\title{Beyond traditional box-covering: Determining the fractal dimension of complex networks using a fixed number of boxes of flexible diameter}
		
		
		
		\author[label1]{Micha\l{} \L{}epek\corref{cor1}}
		\ead{michal.lepek@pw.edu.pl}
		\cortext[cor1]{corresponding author}
		
		\author[label1]{Kordian Makulski}
		
		\author[label1]{Agata Fronczak}
		
		\author[label1]{Piotr Fronczak}
		
		\affiliation[label1]{organization={Faculty of Physics, Warsaw University of Technology},
			addressline={Koszykowa 75}, 
			city={Warsaw},
			postcode={PL-00-662}, 
			country={Poland}}
		
		\begin{abstract}
			In this article, we present a novel box-covering algorithm for analyzing the fractal properties of complex networks. Unlike traditional algorithms that impose a predetermined box size, our approach assigns nodes to boxes identified by their nearest local hubs without enforcing rigid distance constraints. This flexibility leads to a key methodological shift: instead of fixing the box size in advance, we first determine the number of boxes and then compute their average size. We argue that this procedure is fully consistent with the recently proposed scaling theory of fractal complex networks and closely related to the concept of hidden metric spaces in which network nodes are embedded. We demonstrate that our approach not only significantly reduces computational complexity compared to existing methods, but also—despite relaxing constraints on box diameter—covers networks using boxes of more similar sizes than, for instance, the classical greedy coloring (GC) algorithm. To evaluate the effectiveness of our method, we analyze nine complex networks—three model-based and six real-world—representing a broad spectrum: from networks with confirmed fractality, through those with initially uncertain, but here confirmed, fractal properties (such as the Internet at the level of autonomous systems), to large-scale networks that have so far remained beyond the reach of existing algorithms due to their size.
		\end{abstract}
		
		
		
		
	\end{frontmatter}
	
		
\section{Introduction}
		
In this paper, we propose a new method for estimating the box-counting dimension of fractal complex networks \cite{Song_2005, Song_2006, Gallos_2007} and compare it with the currently most widely used technique based on greedy coloring (GC) \cite{Song_2007, Wen_2021, Molontay_2021}. Just as the well-known method for computing the fractal dimension of classical fractals embedded in Euclidean space operationalizes the concept of geometric self-similarity, our approach is grounded in the \textit{generalized} notion of geometric self-similarity—an idea introduced and supported in the recent paper on scaling theory of fractal complex networks \cite{Fronczak_2024}.
		
To rationalize and justify our approach, we begin by recalling the fundamental properties of classical fractals \cite{Feder_1988_book, Rosenberg_2020_book} and explaining how these differ from the structural properties of fractal complex networks \cite{Makulski_2025}. We first introduce the concept of geometric self-similarity and the standard procedures for estimating fractal dimension in classical systems \cite{Bunde_2009_book, Sornette_2006_book}. We then discuss the generalized notion of geometric self-similarity in fractal networks, which led to a consistent scaling theory \cite{Fronczak_2024}. Among the several scaling relations derived within that theory, we highlight those which—together with other well-established concepts such as hub repulsion \cite{Song_2006, Yook_2005} and hidden metric spaces \cite{Serrano_2008, Krioukov_2010, Serrano_2021_book}—served as theoretical guidelines in the development of our box-covering algorithm, which is the central result of this paper.
		
In later sections, we demonstrate that our algorithm not only significantly outperforms the GC method in terms of computational complexity, but also yields objectively more efficient coverings.
		
The core idea behind our algorithm is based on the so-called graph Voronoi diagrams, a geometric partitioning technique introduced in \cite{Erwig_2000} and already adopted for community detection in complex networks \cite{Deritei_2014}. Given the conceptual link between Voronoi tessellations and underlying metric spaces, the strong performance of our algorithm further supports the hypothesis that fractality in networks may emerge from latent geometric structure \cite{Boguna_2021}. This connection opens up several promising directions for future research, which we discuss in the final section of the paper.

The remainder of this paper is organized as follows. In Section~\ref{SecTheory}, we briefly discuss the theoretical background behind the classical box-covering method and highlight the motivation for inverting its logic by fixing the number of boxes (FNB) instead of their diameter in fractal complex networks. Readers primarily interested in the algorithm itself, its implementation, and its performance can skip this section without loss of continuity. In Section~\ref{SecAlgorithm}, we describe the core implementation of the FNB algorithm step by step. Section~\ref{SecData} provides a detailed overview of the fractal complex networks, both model-based and real-world, used for testing the algorithm. In Section~\ref{SecResults}, we present the results of applying the algorithm to estimate the box dimension of fractal networks and assess the quality of the obtained coverings. In Section~\ref{SecVaria}, we explore several possible extensions of the algorithm and introduce measures for evaluating its efficiency. Section~\ref{SecGC} is devoted to comparing the FNB approach with the classical GC algorithm. Finally, Section~\ref{SecFinal} summarizes our main findings and discusses potential directions for further development and applications of the FNB methodology. The article also includes Supplementary Materials (SM) with additional charts and analyses referenced throughout the paper.

\section{Conceptual and theoretical foundations\\of the algorithm}\label{SecTheory}
		
\subsection{Geometric self-similarity of classical fractals}

A defining property of classical fractals—such as the Sierpinski gasket or critical percolation cluster—is geometric self-similarity \cite{Bunde_2009_book}, meaning that their structure appears similar across different scales. This notion can be formalized by considering how a measurable property of the object, such as mass, length, or area, denoted $m(L)$ changes with the observation scale $L$. Specifically, if the object of size $L$ is rescaled by a factor $\lambda>0$, so that the new observation scale becomes $L'$, the quantity $m(L)$ transforms according to the scale-invariant law \cite{Sornette_2006_book}:
\begin{equation}
	m(L') = \lambda^d\, m(L),\quad\text{where}\quad L'=\lambda L, 
\end{equation}
and $d$ is the \textit{fractal (similarity) dimension}. This implies a power-law dependence of the observable:
\begin{equation}\label{mL0}
	m(L) \sim L^d,
\end{equation}
indicating that the measured quantity increases with the scale of observation. 
		
Observing the object at a larger scale (for $\lambda>1$) yields a coarser resolution and a proportionally larger total measure, as encoded by $m(\lambda L)$. In contrast, observing the object at finer resolution (for $\lambda<1$ reveals additional structural detail. As smaller features become accessible, the effective complexity increases—reflected in the growing number of smaller-scale components needed to represent the object. This increasing complexity at finer scales provides the foundation for various methods of quantifying fractality, including the cluster-growing and box-counting approaches, which operationalize the scaling principle described above.
		
In particular, the cluster-growing method, which is typically used for classical fractals embedded in Euclidean space, involves measuring the average mass $\langle m(\ell_B)\rangle$ contained within a distance $\ell_B$ from a designated seed element. If this mass scales as a power law with distance, i.e.,
\begin{equation}\label{df0}
	\langle m(\ell_B) \rangle \sim \ell_B^{\,d_f},
\end{equation}
then the above, empirically obtained scaling relation yields a direct estimate of the so-called \textit{spreading} or \textit{growth} fractal dimension $d_f$, which, for such systems, coincides with the fractal dimension of the object: $d_f=d$.
		
A complementary and widely used method is box-counting. In this approach, the object is covered by a minimal number $N_B(\ell_B)$ of non-overlapping boxes of side length $\ell_B$. Assuming that the content measured within a single box scales as $m(\ell_B)\sim\ell_B^d$, cf. Eq.~(\ref{mL0}), it follows that the total number of boxes required to cover the object scales as
\begin{equation}\label{dB0}
	N_B(\ell_B) \sim \ell_B^{-d_B},
\end{equation}
where $d_B=d$ is the \textit{box-counting} dimension. This relation reflects the assumption that the total size (or mass) of the object remains constant across scales, i.e., $N_B(\ell_B)\cdot m(\ell_B)=\text{const}$. Thus, the growing number of boxes required at finer resolutions quantifies the object's geometric complexity and provides an operational estimate of its fractal dimension.
		
\subsection{Generalized geometric self-similarity\\of fractal complex networks}\label{SecIntroSF}
		
In classical fractals—those embedded in Euclidean space—the similarity dimension $d$, the spreading dimension $d_f$, and the box-counting dimension $d_B$ are all equal. In fractal complex networks, the situation is markedly different: while these networks possess a well-defined box-counting dimension $d_B$~(\ref{dB0})—which, as shown in \cite{Fronczak_2024}, is equivalent to the similarity dimension $d$, and which we draw upon in the present analysis—the concept of a spreading dimension, understood in terms of Eq.~(\ref{df0}), is not properly defined~\cite{Song_2005}. Understanding the nature of this discrepancy—specifically, how geometric self-similarity manifests differently in complex networks—provides a key insight that informs the development of a more efficient box-covering strategy, as discussed later in this work.
		
To clarify this point, let us briefly revisit how the box-counting dimension is defined in complex networks \cite{Song_2005}. It can be estimated by covering the network with boxes understood as non-overlapping sets of nodes, such that the pairwise distances between all nodes within a box do not exceed a predefined diameter $\ell_B$. When such a network undergoes a renormalization procedure—whereby each box is contracted into a single node, and links between boxes are inherited from the original network—the resulting coarse-grained network retains the same structural features: it remains both scale-free and fractal, with the same characteristic exponents.
		
This invariance under renormalization means that, at the global level, fractal complex networks exhibit geometric self-similarity which satisfies the classical scale-invariance law as given by Eq.~(\ref{mL0}). To see this, it suffices to reinterpret the quantity $m(L)$ in (\ref{mL0}) as the number of nodes $N(L)=N$ in the network of diameter $L$. Upon renormalization, the scale transformation becomes $\lambda=L'/L=\ell_B$, and since the size of the renormalized network is given by the number of boxes $N'(L')=N_B(\ell_B)$, one recovers—via substitution—the well-known scaling relation that underpins the box-counting method for networks: $N_B(\ell_B)=N \ell_B^{-d_B}$, cf.~Eq.~(\ref{dB0}).
		
The difficulty in defining a spreading dimension in fractal complex networks stems from the fact that classical self-similarity—based solely on spatial scaling—is insufficient to fully capture the structural heterogeneity of such systems. To address this limitation, a generalized framework of geometric self-similarity has been proposed in Ref.~\cite{Fronczak_2024}, in which the classical scale-invariance relation (\ref{mL0}) is extended by introducing an additional structural parameter. Specifically, the mass function of a box—equal to the number of nodes it contains—is allowed to depend not only on the observation scale $L$, but also on a second variable $k$, which characterizes a relevant local property of the system. In the case of complex networks, this property is taken to be the degree of the most connected node (i.e., the local hub) within a region of diameter $L$. This leads to a generalized scale-invariance law of the form  
\begin{equation}\label{gss0}
	m(L', k') = \lambda^d\, m(L, k), \quad \text{where} \quad L' = \lambda L,
\end{equation}
which, from a mathematical standpoint, defines a class of generalized homogeneous functions  
\begin{equation}\label{mLk0}
	m(L, k) \sim L^\alpha k^\beta.
\end{equation}
This functional form imposes scaling constraints between the microscopic exponents $\alpha$ and $\beta$, and the global similarity dimension $d$, which—as demonstrated in the previous paragraph—is equivalent to the box-counting dimension $d_B$. 

In particular, the generalized self-similarity relation—Eq.~(\ref{gss0})—makes it possible to derive a key scaling law that links the global box-counting dimension to local self-similarity properties of the network:
\begin{equation}\label{dB1}
	d_B=\alpha+\beta d_k,
\end{equation}
where the exponent $d_k$ describes how the degree of a local hub transforms under the renormalization procedure associated with a change in observation scale:
\begin{equation}\label{dk0}
	k' = \lambda^{d_k} k.
\end{equation}
This framework also establishes explicit links between the microscopic exponents $\alpha$ and $\beta$ (\ref{mLk0}), which describe local structure of fractal complex networks, and the macroscopic scaling exponents $d_B$, $\gamma$ (\ref{Pk0}), and $\delta$ (\ref{Pm0}), which characterize their global properties:
\begin{equation}\label{ab0}
	\alpha = \frac{\delta - 2}{\delta - 1} d_B, \quad \text{and} \quad \beta = \frac{\gamma - 1}{\delta - 1}.
\end{equation}
These relations, together with the scaling law—Eq.~(\ref{mLk0})—were previously used in~\cite{Fronczak_2024, Makulski_2025} to validate the generalized self-similarity framework. In Sec.~\ref{SecValidation}, we apply the same reasoning to examine the theoretical soundness of the box-covering algorithm introduced in this work.

Let us now turn to a less obvious, yet highly important consequence of this framework. The strength of the generalized self-similarity approach lies not only in its ability to capture key structural features of fractal networks, but also in its explanatory power: it provides a theoretical basis for understanding why classical measures—such as the spreading dimension $d_f$, defined via mass accumulation from randomly selected seed nodes—fail in complex networks. In what follows, a quantitative explanation for this failure is given, linking it to the heterogeneity and heavy-tailed mass distributions characteristic of fractal, scale-free systems.
		
Specifically, in fractal network with scale-free node degree distribution: 
\begin{equation}\label{Pk0}
	P(k)\sim k^{-\gamma},
\end{equation}
especially for $\gamma=1+d_B/d_k <3$, mass accumulation is highly uneven across the system and strongly influenced by local topology. As shown in \cite{Fronczak_2024}, this results in a broad, heavy-tailed distribution of box masses:
\begin{equation}\label{Pm0}
	P(m)\sim m^{-\delta},
\end{equation}
with $\delta=1+d_B/(d_k\beta)<\gamma$ for $\beta>1$, which undermines the validity of ensemble-averaged measurements and renders the classical spreading dimension operationally ambiguous in such settings. 
		
This theoretical insight leads to important conclusions that can be used to formulate a new, efficient algorithm for estimating the fractal dimension of complex networks—an approach that, as we show below, effectively combines key ideas from both the box-covering and cluster-growing methods.
		
\subsection{Hub-hub repulsion, hidden metric spaces,\\and graph Voronoi diagrams} \label{SecIntro3}

\begin{figure}[t]
	\centering
	\includegraphics[width=0.9\columnwidth]{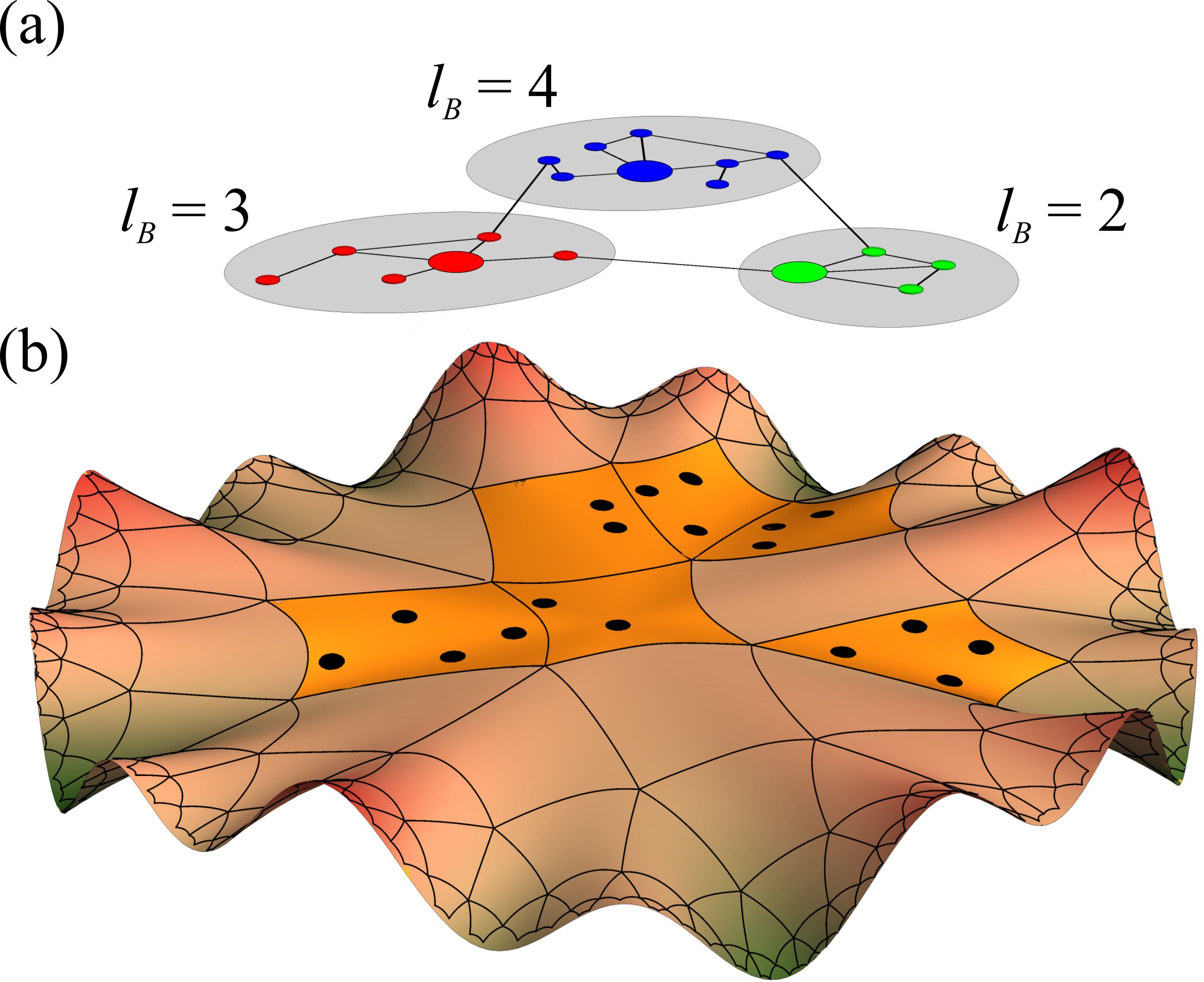}
	\caption{\textbf{Illustration of how the concept of hidden metric spaces promotes hub-hub repulsion in fractal complex networks}. (a)~\textit{Network topology}: Nodes with the highest degrees are enlarged relative to others and serve as the centers (seeds) of graph Voronoi cells (shaded regions, with matching node colors). Note that Voronoi boxes do not have a fixed diameter—nodes equidistant to multiple seeds are assigned randomly to one of the corresponding cells. (b)~\textit{Geometry of the hidden metric space}: Proximity in this latent space influences the likelihood of connections in the observable network—the smaller the distance between nodes, the more similar they are, and the more likely they are to be connected. As explained in the main text, our algorithm is based on the hypothesis that node positions in this hidden space determine local connectivity patterns in the observable network. Accordingly, the graph Voronoi tessellation can be interpreted as an emergent analog of classical box-covering over nodes embedded in an underlying metric space.}
	\label{fig1new}
\end{figure}
		
One of the most intriguing structural features of fractal complex networks is the tendency for high-degree nodes (hubs) to avoid direct connections with one another—a phenomenon known as \textit{hub-hub repulsion}~\cite{Song_2006, Yook_2005}. This behavior contrasts sharply with that observed in typical scale-free networks, where hubs are often densely interconnected. In fractal networks, however, hubs are typically separated by chains of lower-degree nodes, resulting in a more modular and hierarchically nested topology. This organization promotes geometric self-similarity and is regarded as a key condition for the emergence of fractality in networks.
		
Interestingly, this separation of hubs is not merely a topological artifact, but emerges naturally from the generalized notion of geometric self-similarity discussed earlier. In particular, the scaling relation \( k' = \lambda^{d_k} k \), introduced in Eq.~(\ref{dk0}), describes how the degree of a local hub transforms under a change of observational scale. For \( \lambda < 1 \), this relation implies that boxes containing smaller hubs may become indistinguishable at coarser resolutions, while only those with large hubs remain detectable. Since the degree distribution \( P(k) \sim k^{-\gamma} \) is preserved under renormalization~\cite{Song_2005, Fronczak_2024}, the spatial separation of hubs becomes a necessary condition: hubs must be well-separated to maintain the global structure. Thus, hub-hub repulsion can be seen as a geometric manifestation of scale invariance at both global and local levels.
		
This interpretation links naturally to the concept of \textit{hidden metric spaces} \cite{Serrano_2008, Krioukov_2010, Serrano_2021_book}, which posits that network nodes are embedded in a latent geometric space, where the probability of forming a connection decreases with distance (see Fig.~\ref{fig1new}). This  concept has found increasing empirical support in real-world systems. It offers geometric explanations for phenomena such as strong clustering and scale-free degree distributions. While a direct connection between hidden spaces and network fractality has not yet been fully established (as noted in~\cite{Boguna_2021}), the fact that many networks exhibiting clear geometric embeddings also display fractal scaling (cf.~real-world networks analyzed in this study with networks discussed in \cite{Serrano_2021_book}) lends weight to the idea that these two features are closely related. Establishing such a link remains one of the major open questions in the geometry of networks.
		
The relevance of hidden geometry has inspired geometric methods of network partitioning—among them, \textit{graph Voronoi diagrams} \cite{Erwig_2000, Deritei_2014}. By analogy to classical Voronoi tessellations in Euclidean space, graph Voronoi diagrams partition the network into cells centered on selected seed nodes, such that each node is assigned to the seed to which it is closest (in terms of shortest-path distance). In the context of hub-hub repulsion and generalized self-similarity, hubs are natural candidates for seeds, as they tend to dominate the local structure. Moreover, the larger the diameter of a Voronoi cell, the larger the degree of its seed should be, in order to preserve geometric coherence across scales. Importantly, the very idea of Voronoi partitioning implicitly assumes the existence of an underlying distance structure—akin to a hidden metric space (see Fig.~\ref{fig1new}). As shown in the following sections, the strong performance of our box-covering algorithm, which is based on graph Voronoi diagrams, provides operational evidence that such a space may exist and plays a fundamental role in the fractal organization of complex networks.

\begin{table*}[t]
	\footnotesize
	\begin{center}
		\begin{tabular}{ | m{7em} | c | c | c | c | c | }
			\hline
			Network & $N$ & $\langle k \rangle $ & $d$ & $d_B$ (FNB) & $d_B$ (GC) \\
			\hline 
			\hline
			\texttt{SHM} & 1,343,694 & 2.5 & 6,561 & 1.62 (1.63) & { 1.55*} \\
			\hline
			\texttt{(u,v)flowers} & 4,804,002 & 2.4 & 8747 & 1.75 (1.77) &{ 1.67*} \\  
			\hline
			\texttt{nested BA} & 1,000,000 & 3.8 & 62 & 3.85 & 3.62* \\
			\hline
			\texttt{DBLP} & 8,805 & 2.6 & 71 & 2.15 & 2.00 \\
			\hline
			\texttt{brain} & 8,458 & 6.1 & 64 & 2.25 & 2.15 \\
			\hline
			\texttt{proteins} & 4,164 & 21.3 & 32 & 2.10 & 2.37 \\
			\hline			
			\texttt{AS} & 15,488 & 44.5 & 17 & 3.83 & 3.16 \\
			\hline
			\texttt{website} & 2,107,689 & 16.1 & 24 & 7.90 & -  \\
			\hline
			\texttt{road} & 126,146 & 2.6 & 617 & 2.03 & 2.03 \\
			\hline
		\end{tabular}
		\caption{\textbf{Values of the parameters of the fractal networks used in the study}. In the table, $N$ is the number of nodes in the analyzed network, $\langle k \rangle$ is the average node degree, $d$ corresponds to the diameter of the network, and $d_B$ is the fractal dimension obtained using our FNB algorithm and the Song's GC algorithm. The fitting procedure was carried out within the same intervals, marked by the red line in Fig. \ref{fig2new}. Numbers in brackets give theoretical values, if known. The $d_B$ values marked with an asterisk were calculated for networks two orders of magnitude smaller than those given in the table, because the GC algorithm was unable to process networks of the original sizes. For the same reason we were unable to calculate $d_B$(GC) for the \texttt{website} network.\label{table_networks}}
	\end{center}
\end{table*}

\begin{figure*}[t]
	\centering
	\includegraphics[width=0.90\textwidth]{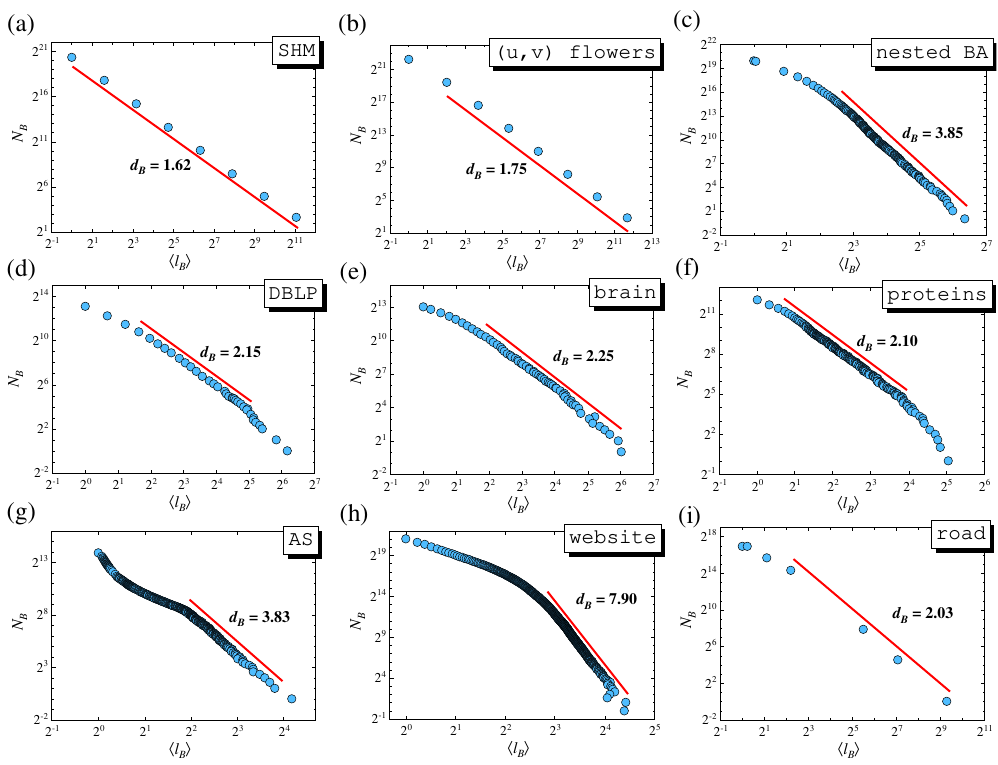}
	\caption{\textbf{Determining the box dimension of fractal networks using the FNB algorithm.}
		Each panel shows a log-log plot of the number of boxes, $N_B$, versus the average box diameter, $\langle\ell_B\rangle$, illustrating the fractal nature of the studied networks according to the scaling relation~(\ref{dB0}). The box-counting dimension values indicated in the figure and listed in Table~\ref{table_networks} were obtained by linear fitting in the intervals marked with red lines (the data fit range has been marked in a similar manner throughout the paper). Further details are provided in the main text.} 
	\label{fig2new}
\end{figure*}

\section{Algorithm}\label{SecAlgorithm}

Our algorithm to find the box-counting fractal dimension of complex network is as follows:
\begin{enumerate}
	\item Choose $k_{cut}$, i.e. the threshold value of the node degree. The nodes of degrees $k_i \ge k_{cut}$ are chosen as local hubs, serving as the centers of the boxes (later, we will discuss alternative methods for selecting central nodes). The highest value of $k_{cut}$ corresponds to the maximum degree of the graph. The number of boxes is the number of hubs:
	\begin{equation}\label{NBfix}
		N_B=\sum_{k_i\ge k_{cut}}N(k_i). 
	\end{equation}
	\item For each node in the network, find its nearest hub. Assign the node to this hub. In case of a tie, where the node is equidistant from two or more hubs, one of them is chosen at random. The hub and its assigned nodes constitute the box.
	\item Determine the diameter $d_{i}$ of each box $i$ by finding the maximal shortest path between any two nodes in the box.
	\item Calculate $\langle\ell_B\rangle$ as an average size of the box in the network
	\begin{equation}
	\langle \ell_B\rangle=1+\frac{\sum_{i=1}^{N_B} d_i}{N_B}.\label{srlb}
	\end{equation}
	\item Repeat points 1 -- 4 for different $k_{cut}$.
\end{enumerate}

We will refer to this algorithm as a fixed number of boxes algorithm, or FNB in short. In Sec.~\ref{SecVaria}, where we compare the results of two alternative strategies for central node selection — namely, the original hub-based method and the random-seeding variant — we introduce additional labels, FNB-h and FNB-r, to denote the respective versions.

Before moving on, let us highlight a few important implementation details and clarifications regarding the algorithm’s behavior and assumptions. First, as can be seen from Eq.~(\ref{srlb}), the size of individual boxes is assumed to be: $\ell_B=d_i+1$. This convention is in line with the previous approach of Song et al. \cite{Song_2005}. It guarantees that the size of boxes containing individual nodes is non-zero, so that points $(\ell_B,N_B)=(1,N)$, unlike points $(0,N)$, can be shown in log-log plots illustrating the box-counting method~(\ref{dB0}), cf. Fig. \ref{fig2new}.

Second, implementation of the algorithm can be done with the use of the burning (or 'infecting'; realized by breadth-first search) strategy, when we sequentially burn out next nearest neighbors of each hub to find its closest nodes. It guarantees that, in the final step, each node is assigned to some hub, and that at least one path between any two nodes within a box is entirely contained within that box. This ensures that disconnected boxes are not allowed — in contrast to some earlier methods, such as the GC algorithm~\cite{Molontay_2021}.

As supplementary materials, we include two short movies that illustrate this issue using a 30-node subgraph extracted from the \texttt{'road} network, which is one of the six real-world networks analyzed later in this manuscript. \textit{Movie~S1} demonstrates the behavior of our FNB algorithm, while \textit{Movie~S2} shows the behavior of the GC algorithm. This visual comparison highlights the differences in how each method assigns nodes to boxes and handles network connectivity.

Third, the computational complexity of the algorithm needed to find one tuple $(\langle\ell_B\rangle,N_B)$ is $O(N)$.  It is a much more efficient method than the GC algorithm, where it is $O(N^2)$. It allows the analysis of networks two orders of magnitude larger than the GC algorithm. The pseudocode of the FNB algorithm and a detailed analysis of its complexity are provided in the supplementary materials. The implementation of the algorithm in Python is available online \cite{Fronczak_2025}.

\section{Model-based and real-world network data}\label{SecData}

For this study, we tested our algorithm on several model-based and real complex networks.

The model-based networks were:

\begin{itemize}
	\item '\texttt{SHM}' (Song–Havlin–Makse) model \cite{Song_2006}, with parameters: $m=2$, $p=2$, and number of iterations $g=8$.
	\item Deterministic model of fractal complex networks called '\texttt{(u,v)flowers}' \cite{Rozenfeld_2007, Rozenfeld_2009_book}, with parameters: $u=3$, $v=4$, and number of iterations $g=8$.
	\item Evolving model of fractal complex networks called '\texttt{nested BA}' model \cite{Fronczak_2024, Makulski_2025}, with parameters: $n_{\text{max}}=500$, $a=2$, $\tau=2$, $A=0.5$, and size $N=10^6$.
\end{itemize}

The \texttt{SHM} and \texttt{(u,v)flower} networks are examples of deterministic network models. In these models, the network's structure is generated through a predefined, rule-based iterative process, rather than through stochastic connections. As a result, node degrees in these networks take on only a limited set of specific, discrete values determined by the construction rules. Detailed construction procedures for these models are provided in the cited references.

\begin{itemize}
	\item '\texttt{DBLP}' coauthorship network: DBLP is a digital library of article records published in computer science \cite{Tang_2012, DBLP_repo}. In this study, similarly as in Refs.~\cite{Fronczak_2024,Fronczak_2022}, we use the 12th version of the dataset (DBLP-Citation-network V12; released April 2020, which contains information on approximately 4.9 M articles published mostly during the last 20 years). We ourselves processed the raw DBLP data into the form of coauthorship network, from which we extracted the network backbone by imposing a threshold on the minimum number of joint papers ($\ge20$) two scientists should have. This procedure significantly reduced the size of the studied network (from 4.9 M nodes and 12.5 M links to 8.8 k nodes and 11.4 k edges), but thanks to it the network became naturally fractal.
	
	\item Human '\texttt{brain}' network: The network is based on functional magnetic resonance imaging (fMRI). The fMRI data consists of temporal series, known as the blood oxygen level dependent (BOLD) signals, from different brain regions. To build brain networks, the correlations $C_{ij}$ between the BOLD signals are calculated and the two nodes (brain regions) are connected if $C_{ij}$ is greater than some threshold value $T$. In our case, we assumed $T = 0.8$. The brain network analysed here was used in Refs.~\cite{Fronczak_2024, Gallos_2012, Reis_2014} and can be found at \cite{brain_repo}.
	
	\item Human '\texttt{proteins}' interaction network: The network is generated from the STRING - a database of known and predicted protein-protein interactions \cite{string}. Two nodes are connected if predicted association between genes based on observed patterns of simultaneous expression of genes (coexpression) is larger than 0.5. We analyze the largest connected component only.
	
	\item Autonomous systems '\texttt{AS}' network: A network dataset representing the topology of the Internet at the level of autonomous systems (AS). In this network, nodes correspond to individual AS entities —typically Internet service providers, large organizations, or network operators— and edges represent direct peering or routing relationships between them. The dataset is collected and curated by the Center for Applied Internet Data Analysis (CAIDA) and reflects snapshots of the global Internet infrastructure at a given time. This type of network is widely used in studies of Internet topology, network resilience, and the analysis of large-scale communication systems. In this study, we used a dataset from January 1, 2024, in which we included only provider-customer connections \cite{caida_as}. The resulting network contains 15 k nodes and 345 k edges.

	\item Baidu Baike '\texttt{website}': a large network consisting of 2.1 M nodes and 16.1 M edges, representing the structure of hypertext links within the Chinese online encyclopedia Baidu Baike. In this dataset, nodes correspond to articles, and edges represent hyperlinks between them. The dataset is available through the Network Data Repository \cite{niu2011zhishi, Rossi_2015_repo}. To the best of our knowledge, this website is the largest real network whose fractal structure has been studied to date.

	\item \texttt{'Road'} network: A network of continental road connections in the United States \cite{Rossi_2015_repo}. It contains 126 k nodes adn 162 k edges. It stands out from other networks due to its very narrow degree distribution, with node degrees ranging from 1 to 7.
\end{itemize}

\section{Results obtained using the FNB algorithm}\label{SecResults}

\subsection{Box-counting dimension}\label{SecdB}

At the outset, we note that, since our method first determines the number of boxes and then evaluates the corresponding average box size, Eq.~(\ref{dB0}) can be equivalently rewritten as
\begin{equation} \label{eq_lbnb}
	\langle \ell_B \rangle (N_B) \sim N_B^{-1/d_B},
\end{equation}
which, of course, does not change the value of $d_B$, but more accurately reflects our approach—an inverted version of the classical box-counting method. Nevertheless, to enable direct comparison with earlier studies, we continue to use the conventional form of data presentation, $N_B(\langle \ell_B \rangle)$.

This seemingly minor technical adjustment underscores a broader methodological shift introduced by our approach and leads to one of the most important advantages of the FNB algorithm. Specifically, it allows the box size $\langle \ell_B \rangle$ to be treated as a continuous variable ($\langle \ell_B \rangle \in \mathbb{R}_+$), rather than restricting it to discrete values ($\ell_B \in \mathbb{N}$). This enhancement enables much higher resolution in the scaling analysis of $N_B(\langle \ell_B \rangle)$, resulting in a larger number of data points and more precise estimates of the fractal dimension. In contrast, earlier methods relied on a small set of integer box sizes, which often led to ambiguous or inconsistent results (cf. Fig.~3 in Ref.~\cite{Zhang_2018}, containing only five data points; Fig.~3(g) in Ref.~\cite{Kim_2007}; and Fig.~\ref{fig7new}(e,f) in Sec.~\ref{SecGC} of this article).

\begin{figure*}[t]
	\centering
	\includegraphics[width=0.9\textwidth]{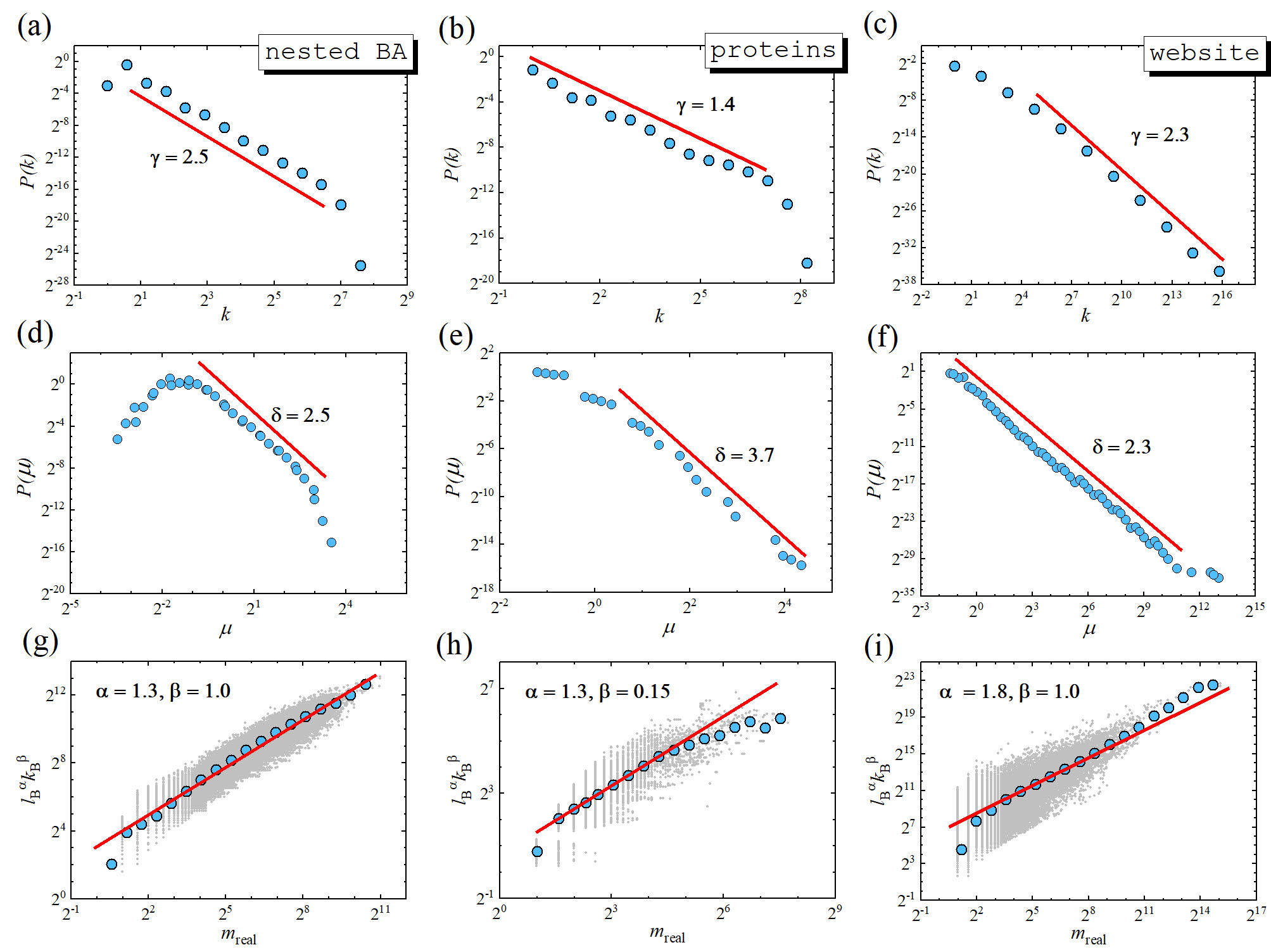}
	\caption{\textbf{Theoretical validation of the FNB algorithm}. The graphs placed in the same column refer to the same network (i.e. \texttt{nested BA}, \texttt{proteins}, and \texttt{website}) and those placed in the same row to the same scaling relation. And so, respectively: The graphs (a--c) in the top row show logarithmically averaged node degree distributions $P(k)$ (\ref{Pk0}) on a double logarithmic scale. The graphs (d--f) in the middle row show, in an analogous way, the box mass distributions, $P(\mu)$ (\ref{Pm0}), where $\mu=m/\langle m\rangle$, where $\langle m\rangle=N/N_B=\langle\ell_B\rangle^{d_B}$ (\ref{dB0}). The meaning of red lines in the graphs (a--f) is the same as in Fig.~\ref{fig2new}. Finally: The graphs (g--i) in the bottom row present scaling of the masses of boxes according to Eq.~(\ref{mLk0}), as described in the main text. The red lines in the graphs (g--i) of slope one stand for pure proportionality to highlight the agreement between the theoretical predictions and the data obtained from real networks.}
	\label{fig3new}
\end{figure*}

In Fig.~\ref{fig2new}, we present the results of the analysis of both model-based and real-world fractal networks. These networks, along with their key parameters, are listed in Table~\ref{table_networks}, which also reports the empirical values of their box-counting dimensions obtained using our FNB algorithm and the classical GC method.

In the context of nearly two decades of research on fractal complex networks, the results shown in Fig.~\ref{fig2new} can be grouped into several meaningful categories, each highlighting different strengths of our method. First, the algorithm correctly identifies as fractal those networks whose fractal nature has already been confirmed in earlier studies using other methods, such as the GC algorithm—this serves as a baseline validation of our approach. This category includes both synthetic networks and real-world examples such as the \texttt{DBLP} and \texttt{brain} networks~\cite{Fronczak_2024}. More importantly, however, the algorithm reveals clear fractal scaling in networks whose fractal status has remained uncertain. This applies in particular to the \texttt{proteins} and \texttt{AS} networks, for which previous analyses yielded inconclusive or inconsistent results. In addition, the figure shows results for two networks that do not belong to any of the above categories. The first is the \texttt{website} network, which—due to its enormous size—has never been subjected to fractal analysis. The second is the \texttt{road} network, which, although not scale-free in terms of degree distribution, is known to exhibit fractal features and is included here to demonstrate the broader applicability and versatility of our method.

Based on these results, we conclude that the FNB algorithm produces robust and interpretable outputs across a wide spectrum of network types, from well-established fractal networks to previously ambiguous cases. Importantly, all data presented in Fig.~\ref{fig2new} were obtained from a single execution of the algorithm per network. To verify the stability of the results with respect to the algorithm's inherent randomness—arising from the random assignment of nodes in the case of ties—we performed multiple independent runs for all analyzed networks. In all cases, we found no significant variation in the resulting fractal dimension~$d_B$.

\subsection{Theoretical validation of the FNB algorithm}\label{SecValidation}

While the primary goal of any box-covering algorithm is to determine the box-counting dimension $d_B$ of a given network, an equally important but often overlooked question is whether the identified boxes themselves reflect meaningful, scale-invariant structure. Ideally, such an algorithm should not only yield the expected scaling $N_B(\ell_B) \sim \ell_B^{-d_B}$, but also group nodes into self-similar boxes consistent with the theoretical framework of fractal networks. This issue is especially relevant given that the scaling function $N_B(\ell_B)$ provides no direct information about the internal consistency or interpretability of the boxes themselves. Fortunately, the framework of generalized self-similarity offers a theoretical toolset for such a validation (see Sec.~\ref{SecIntroSF}). 

Readers familiar with the study of modularity in networks may find the analogy helpful: just as community detection algorithms can be tested against benchmark graphs with known modular structure \cite{Lancichinetti_2008, Fronczak_2013} using dedicated comparison metrics~\cite{Deritei_2014}, so too can box-covering algorithms be assessed based on whether the structures they extract obey well-defined scaling relations. In the case of fractal networks, these relations are derived from established theory and provide a principled basis for such evaluation.

To carry out this theoretical validation, we adopt a method previously developed in our own studies. In~\cite{Fronczak_2024}, we used it to assess whether synthetic and empirical fractal networks conform to the concept of generalized self-similarity, applying the classic GC algorithm for box covering. Later, in~\cite{Makulski_2025}, we applied the same framework to evaluate a generative model of fractal networks, testing whether it correctly reproduced the internal organization of self-similar boxes. In the present work, we use this methodology to assess the quality of the box assignments produced by the FNB algorithm.

The core idea behind the method is to compute the microscopic scaling exponents $\alpha$ and $\beta$—see Eqs.~(\ref{ab0})—based on measurable macroscopic exponents: the box-counting dimension $d_B$, the degree distribution exponent $\gamma$, and the box mass exponent $\delta$—see Figs.~\ref{fig2new}(c,f,h) and~\ref{fig3new}(a--f). These values of $\alpha$ and $\beta$ are then used to test whether the masses of self-similar boxes—identified by the algorithm—are consistent with the theoretical relation linking box mass $m$ to box diameter $\ell_B$ and hub degree $k_B$, as given by Eq.~(\ref{mLk0}). When this consistency is observed, it indicates that the algorithm not only recovers the global fractal properties of the network but also correctly identifies its internal self-similar structure. The graphs in the bottom row of Fig.~\ref{fig3new} (see also Figs. S1 and S2 in Supplementary Materials) show that the consistency test yields very good results for our algorithm.

\section{Variants of the FNB algorithm}\label{SecVaria}

\begin{figure*}[t]
	\centering
	\includegraphics[width=0.90\textwidth]{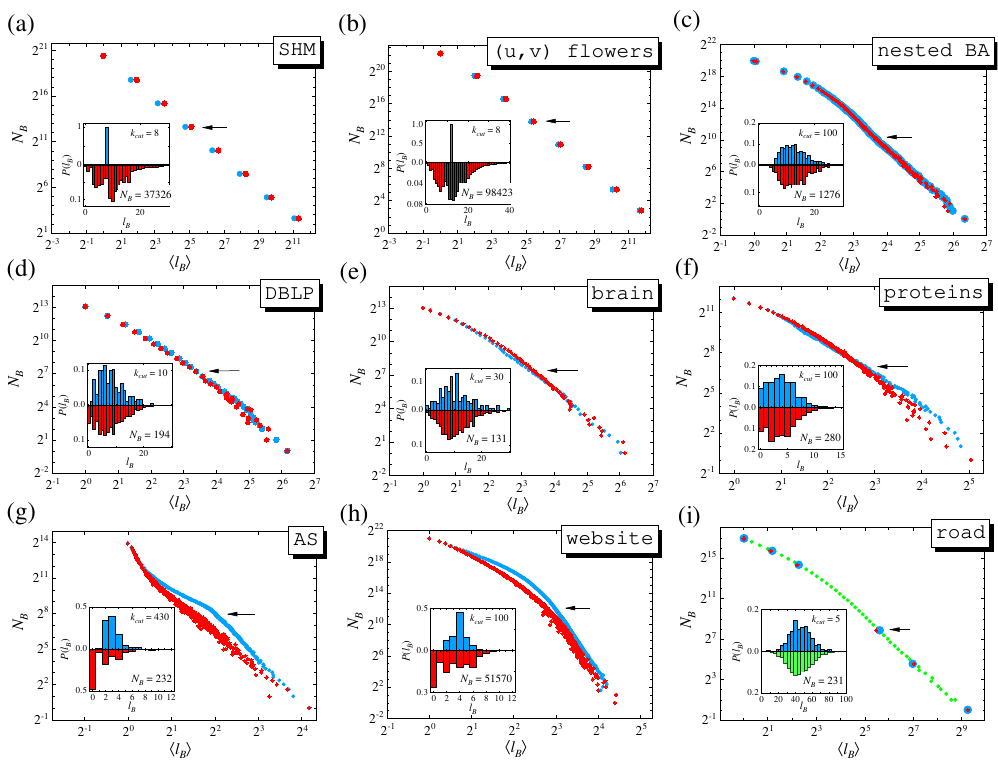}
	\caption{\textbf{Comparison of the scaling functions $N_B(\langle\ell_B\rangle)$ obtained using two variants of the FNB algorithm}. The blue data series show the original FNB-h algorithm, the red series shows the FNB-r algorithm with random seeding, but for the same number of boxes as FNB-h, while the green series shows FNB-r with any number of boxes. Additional small graphs inserted on each graph show the distribution of box diameters $P(\ell_B)$ obtained for a fixed number of boxes marked on the main graph with a black arrow.}
	\label{fig4new}
\end{figure*}

In the original formulation of the FNB algorithm (see Sec.~\ref{SecAlgorithm}), nodes with the highest degree are selected as the initial seeds for each box. This choice is motivated by the central role that hubs typically play in the network’s topology — they naturally serve as anchors around which surrounding nodes can be efficiently grouped (see Sec.~\ref{SecIntro3}). However, one may naturally question whether alternative seed-selection strategies — for example, random selection or selections based on other centrality measures — could lead to comparable or even improved box-covering efficiency and quality.

\begin{figure*}
	\centering
	\includegraphics[width=0.75\textwidth]{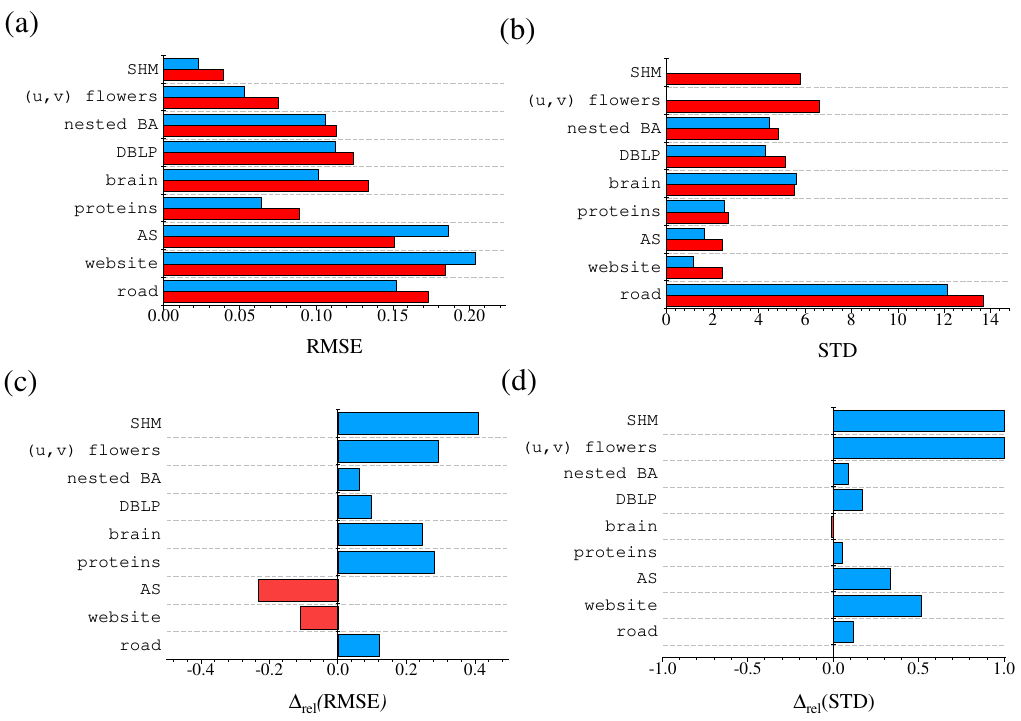}
	\caption{\textbf{Comparison of the two FNB algorithm variants using two quantitative metrics}: the root-mean-square error (RMSE) of the linear fit to the complete $N_B(\langle \ell_B \rangle)$ dataset, and the standard deviation (STD) of the box diameter distribution $P(\ell_B)$. Panels (a) and (b) show the absolute values of both metrics across all networks, while panels (c) and (d) present the relative differences between the FNB-h and FNB-r seeding strategies. Blue bars correspond to the original hub-based strategy (FNB-h), and red bars to the random-seed variant (FNB-r). In panels (c) and (d), the color of each bar indicates which strategy performs better according to the respective metric.}
	\label{fig5new}
\end{figure*}

To investigate this, we conducted a comparative analysis of two variants of the FNB algorithm: the original version with highest-degree nodes as seeds (FNB-h) and a randomized version where the initial nodes are selected uniformly at random (FNB-r). A direct comparison of both strategies across multiple network types is shown in Fig.~\ref{fig4new}. In both cases, the number of boxes $N_B$ was fixed for each configuration according to Eq.~(\ref{NBfix}). The only exception was the \texttt{road} network, where — due to the limited number of available points on the $N_B(\langle\ell_B\rangle)$ graph for FNB-h (see Fig.~\ref{fig2new}(i)) — we extended the analysis for FNB-r to include an arbitrary number of randomly selected seeds.

Before assessing which variant is more effective, it is important to clarify what "efficiency" means in the context of our algorithm. In standard box-covering approaches, efficiency is typically measured by the number of boxes of fixed diameter $\ell_B$ required to cover the network — the fewer, the better. In our case, however, the number of boxes is fixed in advance, and their diameters are allowed to vary. This inversion of the standard logic renders the conventional efficiency criterion inapplicable.

Although, for most networks in our dataset (see Fig.~\ref{fig4new}(a–f,i)), the empirical scaling functions $N_B\langle \ell_B \rangle$ are nearly identical for both variants — and even suggest a slight advantage of FNB-r in the \texttt{AS} and \texttt{website} networks — a more careful analysis reveals that this apparent superiority of FNB-r is misleading.

To illustrate this, consider a simple example: a network of $N = 1000$ nodes is partitioned into 100 boxes, where 99 boxes each contain a single isolated node (i.e., zero diameter), while the last box covers the remaining network. The resulting average box diameter would be close to zero — deceptively implying high efficiency — even though the covering is clearly unbalanced and uninformative.

This is precisely the situation we observe in the \texttt{AS} and \texttt{website} networks, where FNB-r tends to generate a large fraction of trivial boxes of diameter zero. While formally valid in terms of coverage, these boxes distort the diameter distribution and artificially lower the mean box size $\langle \ell_B \rangle$, producing a false impression of higher efficiency.

\begin{figure*}[t]
	\centering
	\includegraphics[width=0.65\textwidth]{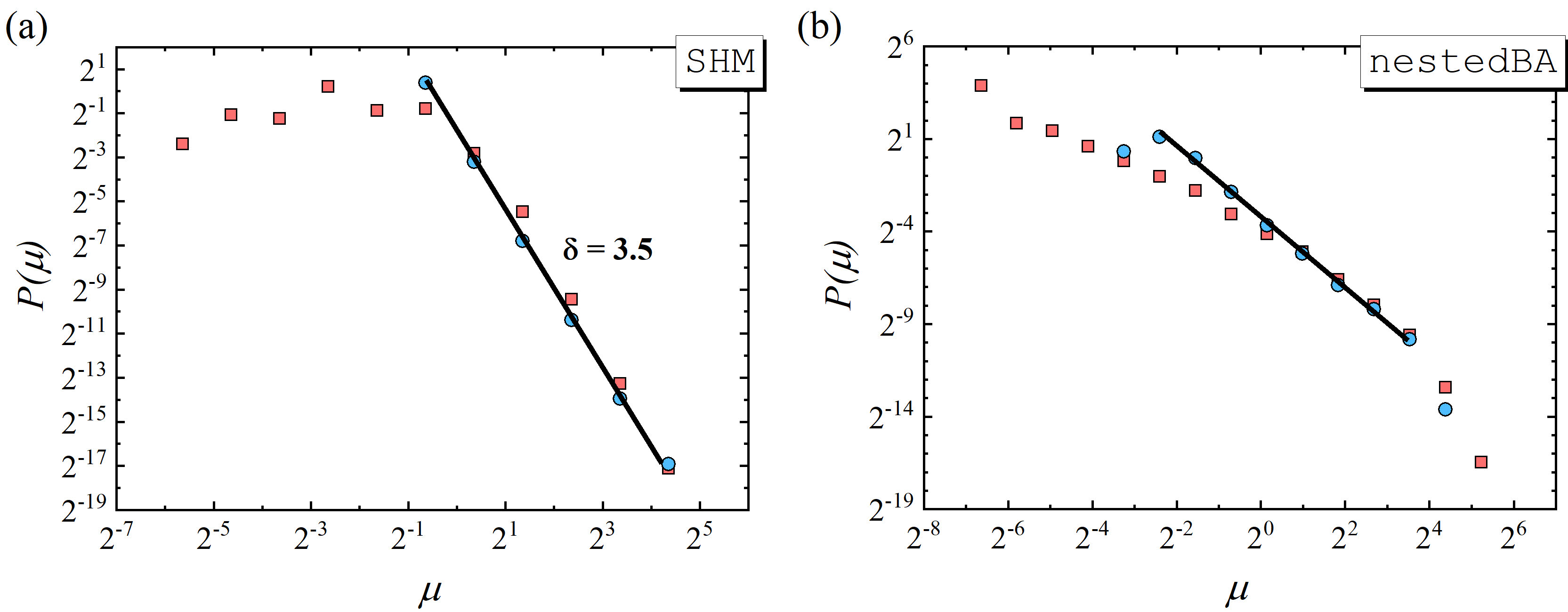}
	\caption{\textbf{Distributions of box masses $P(\mu)$} obtained using FNB-h (blue points) and FNB-r  (red points) for: (a) the \texttt{SHM} network and (b) the \texttt{nested BA} network. Here, a different set of parameters has been used for which the advantage of the original strategy becomes more apparent, $N=10^6, a=1, A=0.2, \tau=2, n_{max}=5000$. The box covering was performed in both cases using $k_{cut}=20$ (FNB-h) and the corresponding number of boxes $N_B$ (FNB-r). } 
	\label{fig6new}
\end{figure*}

To properly evaluate the quality of the coverings produced by the two algorithms, one must go beyond averages and examine the full distributions of box diameters $P(\ell_B)$. To this end, we analyzed $P(\ell_B)$ for a specific number of boxes corresponding to approximately half the vertical range in log-log scale — this value is marked by a black arrow in each panel of Fig.~\ref{fig4new}. These distributions, shown as insets, reveal substantial differences between the two methods, particularly in deterministic networks such as the Song–Havlin–Makse (\texttt{SHM}) model (Fig.~\ref{fig4new}(a)) and the \texttt{(u,v)-flowers} model (Fig.~\ref{fig4new}(b)). In these cases, FNB-h yields unique box diameters (single peaks on the blue histograms), while FNB-r results in a much broader distributions.

To quantify this effect, we introduced an additional efficiency criterion: the dispersion of box diameters around the mean. Ideally, this spread should be as small as possible, indicating that the network is partitioned into boxes of comparable size and avoiding the presence of large or trivial outliers.

To capture this, we used two complementary metrics: the standard deviation (STD) of $P(\ell_B)$ and the root-mean-square error (RMSE) of the linear fit to the full $N_B(\langle \ell_B \rangle)$ dataset. These metrics are compared in Fig.~\ref{fig5new}. More specifically, Fig. 5(a,b) shows the absolute values of these metrics for each network, while Fig. 5(c,d) shows the relative differences between variants, defined as:
\begin{equation}
	\Delta_{\mathrm{rel}}(\mathrm{X})=\frac{\mathrm{X_r}-\mathrm{X_h}}{\mathrm{X_r}},
\end{equation}
with $\mathrm{X}$ being either RMSE or STD.

Positive values of $\Delta_{\mathrm{rel}}(\mathrm{RMSE})$ — blue bars in Fig.~\ref{fig5new}(c) — indicate cases where FNB-h provides a better fit to the scaling function than FNB-r. As expected, the only exceptions are again the \texttt{AS} and \texttt{website} networks, where the RMSE is lower for FNB-r due to the inflation of zero-diameter boxes.

Likewise, positive values of $\Delta_{\mathrm{rel}}(\mathrm{STD})$ — Fig.~\ref{fig5new}(d) — highlight networks where FNB-h yields narrower and more regular diameter distributions — a desirable property if one assumes the existence of an underlying metric space structure.

Finally, Fig.~\ref{fig6new} presents distributions of box masses $P(\mu)$ obtained for both algorithm variants. For the \texttt{SHM} network (Fig.~\ref{fig6new}(a)), where the theoretical exponent $\delta = 3.5$ is known (indicated by the reference line), FNB-h reproduces this scaling, while FNB-r fails to do so. A similar trend is observed for the \texttt{nested BA} network (Fig.~\ref{fig6new}(b)), where no analytical benchmark is available — yet the FNB-h variant clearly identifies a well-defined scaling regime suitable for reliable exponent estimation.

In addition to these two variants, we also explored alternative seeding strategies based on different node centrality measures, including clustering coefficient, average nearest-neighbor degree, and two-step rich-club connectivity. The outcomes of these tests, provided in the Supplementary Materials, confirm that the original hub-degree-based seeding consistently offers the most reliable and efficient results, and we therefore recommend its use for fractal analysis of complex networks.

Given the conceptual similarity between box-covering and community detection, we also tested a non-local clustering approach based on the Leiden algorithm \cite{Traag_2019}, using its \textit{resolution} parameter to control the number of detected modules (corresponding to the number of $N_B$ boxes in FNB algorithm). However, this method did not prove to be more effective or promising for accurate fractal dimension estimation. Therefore, we decided to include the corresponding results only in the Supplementary Materials to avoid distracting from the core message of the paper.

\begin{figure*}[t]
	\centering
	\includegraphics[width=0.90\textwidth]{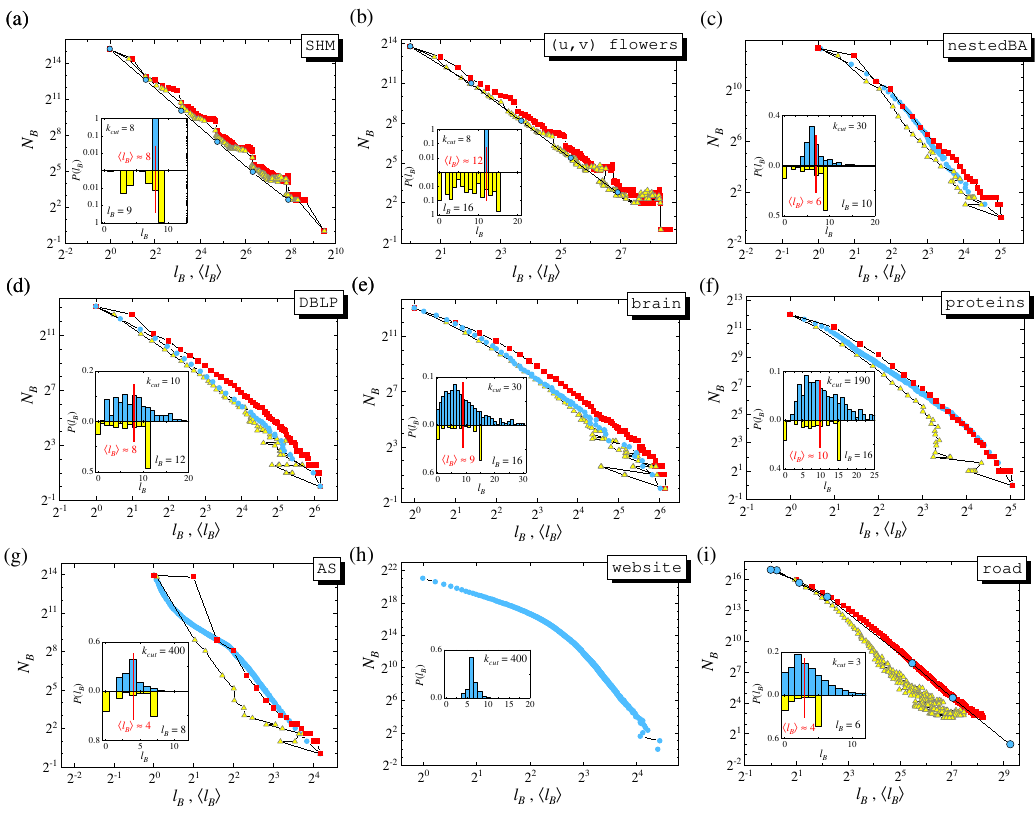}
	\caption{\textbf{Comparison between two box-covering methods: the FNB algorithm and the GC method}. The main panels show the scaling functions — Eq.~(\ref{dB0}) — obtained using both methods. Results for the FNB algorithm are shown as blue circles. Results for the GC algorithm are presented in two ways: red squares correspond to $N_B$ vs. $\ell_B$, and yellow triangles to $N_B$ vs. $\langle\ell_B\rangle$. The insets display the box size distributions $P(\ell_B)$ obtained from coverings with the same $\langle\ell_B\rangle$ (marked by the red vertical lines in inset graphs).} 
	\label{fig7new}
\end{figure*}

\begin{figure*}
	\centering
	\includegraphics[width=0.75\textwidth]{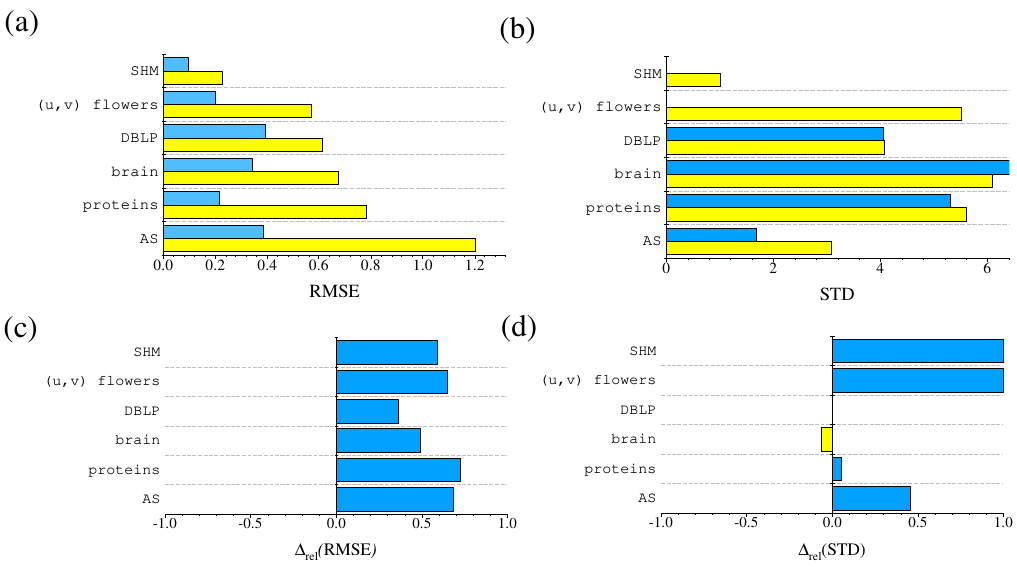}
	\caption{\textbf{Comparison of the FNB and GC algorithms using two quantitative metrics}: the root-mean-square error (RMSE) of the linear fit to the complete $N_B(\langle \ell_B \rangle)$ dataset, and the standard deviation (STD) of the box diameter distribution $P(\ell_B)$. Panels (a) and (b) show the absolute values of both metrics across the analyzed networks (cf.~Fig.~\ref{fig6new}), while panels (c) and (d) present the relative differences between the FNB and GC methods. Blue bars correspond to the FNB algorithm, and yellow bars to the GC method. In panels (c) and (d), the color of each bar indicates which algorithm performs better according to the respective metric.}
	\label{fig8new}
\end{figure*}

\section{Comparison of GC and FNB algorithms}\label{SecGC}

In the following, we will discuss the differences between the proposed algorithm and the previous approach based on greedy coloring (GC). We choose the original box covering algorithm developed by Song et al. \cite{Song_2005} as a reference model as it is the most widely-known box covering algorithm, which was also used for the same purpose (as a benchmark model) in the recent review of box-covering algorithms by Kov\'{a}cs et al.~\cite{Molontay_2021}.

The first distinction lies in how our algorithm generates data points on the $N_B(\langle \ell_B \rangle)$ plot: it produces a data point for each unique node degree observed in the network. As a result, the total number of data points corresponds directly to the number of observable node degrees. In some networks, such as \texttt{(u,v)-flowers} and \text{SHM} models, this characteristic yields a relatively lower number of data points, though it does not compromise the precision of fractal dimension estimation, see Fig.~\ref{fig2new}(a,b). Conversely, in networks with relatively small diameters, like \texttt{protein}, \texttt{website} or \texttt{AS} networks, this feature allows for a higher density of data points, enhancing the interpretability of the plot, particularly in evaluating potential fractal properties, see Fig.~\ref{fig2new}(f,g,h).

Before comparing the two algorithms, it is essential to ensure that we are evaluating corresponding metrics across both outputs. An interesting observation regarding the GC algorithm during box size distribution analysis is the emergence of box sizes that differ from the specified box size $\ell_B$. Although the algorithm initially sets a specific target size, the final box sizes can vary significantly (see yellow distributions of box sizes in the insets of Fig. \ref{fig7new}). This raises an important question regarding the representation of $N_B$: should it be plotted as a function of the target $\ell_B$ or, rather, the average box size $\langle \ell_B \rangle$ — like in our FNB approach. Using the actual average box size $\langle \ell_B \rangle$ offers a more accurate reflection of the distribution. It also facilitates a direct comparison with box-covering methods such as the FNB algorithm, which similarly considers average box sizes. This approach can improve both the precision and interpretability of fractal analysis in network studies.

In Fig.~\ref{fig7new}, we display three data series: two derived using the GC algorithm (with red squares plotted against $\ell_B$ and yellow triangles against $\langle \ell_B\rangle$) and one obtained using the FNB algorithm (blue circles).  For the \texttt{website} network, due to its size, we were unable to apply the GC algorithm. Therefore, in Fig.~\ref{fig7new}, the panel (i) presents only the results obtained with FNB algorithm.

The plots shown in Fig. \ref{fig7new}(a) and \ref{fig7new}(b) suggest that for synthetic fractal networks, both algorithms yield similar results. However, only the results of the linear fit (see Table 1) reveal that the FNB algorithm, with an error below 1\% relative to the theoretical value, clearly outperforms the GC algorithm, which exhibits an error exceeding 5\% for both networks.

Figures \ref{fig7new}(d) and \ref{fig7new}(e) indicate that in these two cases, both algorithms produce comparable results — but only when adopting our proposed way of presenting the GC algorithm’s results, using $\langle \ell_B\rangle$ instead of $\ell_B$ on the horizontal axis (yellow points). Figures \ref{fig7new}(f) and \ref{fig7new}(g) illustrate cases in which the FNB algorithm notably outperforms the GC algorithm by successfully identifying fractality in networks where the GC method either fails or yields inconclusive results. In the case of the protein interaction network, the fractal property is partially obscured by large fluctuations in the lower part of the graph. The \texttt{AS} network is an example of a case where previous studies remained inconclusive due to a limited number of data points — a problem caused by the low network diameter affecting the GC method, as it allows only integer-valued box sizes. In the case of the \texttt{road} network (Fig. \ref{fig7new}(i)), the results of the original GC algorithm plotted against $\ell_B$ outperform those plotted against $\langle \ell_B\rangle$ (in opposite to the previous cases). Nevertheless, the FNB algorithm produces valid results also in this case.

Let us now compare the box size distributions for both algorithms, as shown in the insets of Fig.~\ref{fig7new}. These distributions were generated using $\ell_B$ (for the GC algorithm) and $k_{cut}$ (for the FNB algorithm), set to achieve approximately the same average box size $\langle \ell_B \rangle$ (indicated by the red lines in the insets). While presenting the number of boxes obtained with GC algorithm as a function of average box size may appear reasonable, this average does not accurately reflect the actual distribution of box sizes. For the GC algorithm, the distribution of box sizes is U-shaped (yellow distributions in the insets), and the mean represents one of the least common values. In contrast, the FNB algorithm generates a nearly bell-shaped distribution (blue distributions in the insets), where box sizes near the mean are frequently observed.

We compared both algorithms in a similar manner as we did in the previous section when analyzing variants of the FNB algorithm, i.e. calculating the root-mean-square error (RMSE) of a linear fit to the entire $N_B(\langle \ell_B \rangle)$ dataset, and the standard deviation (STD) of the box diameter distribution, $P(\ell_B)$. The results of these comparisons are shown in Fig. \ref{fig8new}. They confirm our previous visual observations: the FNB algorithm enables obtaining better fractal characteristics (see Fig. \ref{fig8new}(c)) as well as significantly more compact distributions of box diameters (see Fig. \ref{fig8new}(d)). 

In the Supplementary Materials, we also demonstrate that the FNB algorithm’s speed surpasses that of the GC algorithm by several orders of magnitude, enabling the analysis of networks larger than $10^7$ nodes within a reasonable time—something that is not feasible with the GC algorithm.

Finally, it is essential to reiterate that the burning strategy ensures all boxes identified by the FNB algorithm are connected, i.e., at least one path between any two nodes within a box is fully contained within that box. This property does not hold for the GC algorithm, where some boxes may consist of disconnected components (see movie \textit{Movie~S2} in the Supplementary Materials).

\section{Summary and concluding remarks}\label{SecFinal}

We have proposed a new algorithm for detecting fractal scaling in complex networks that offers a conceptual shift from classical box-covering approaches. Instead of fixing a box diameter and counting how many boxes are needed to cover a network, our method fixes the number of boxes and measures their average size. This reversal not only simplifies computation but also makes the method more flexible and more naturally aligned with the intrinsic structure of many real networks. In comparative tests on nine networks of varying origin and complexity, our algorithm not only proved computationally efficient but also yielded box-size distributions that were more homogeneous than those produced by standard methods such as greedy coloring.

What distinguishes our approach is that it implicitly merges two classic methods of fractal analysis known from Euclidean geometry: box-covering and cluster growth. This hybrid character makes the algorithm well-suited not only for complex networks but also for spatial systems such as planar or geometric graphs. In fact, our analysis of the U.S. \texttt{road} network—whose topology is far from that of typical complex networks—suggests that the method can be effectively adapted to a broader class of structures, including classical fractals.

We also believe that the algorithm can be naturally extended to networks with weighted (and even directed) edges, where distances between nodes are not restricted to integers, see e.g.~\cite{Wei_2019}. Since node assignment in our method is based on continuous distance from selected seeds, such generalization requires no fundamental changes to the algorithm’s core. Moreover, the flexibility of our framework opens the door to multifractal analysis, see e.g.~\cite{Liu_2015}, in which a spectrum of scaling exponents is used to capture local heterogeneities in scaling behavior—providing a more detailed and differentiated picture of self-similarity.

Finally, we point to an intriguing direction for further research: the use of graph Voronoi diagrams in conjunction with our method. This line of inquiry connects naturally with the idea of hidden metric spaces—a concept increasingly recognized as central to the geometry of complex networks. Voronoi partitions on graphs are also closely related to community structure \cite{Deritei_2014}, which suggests that fractality and modularity may share deeper conceptual and methodological foundations. For instance, hierarchical community organization resembles geometric self-similarity across scales, while overlapping communities find a natural counterpart in boundary or "ambiguous" nodes that lie at comparable distances from multiple local hubs and thus may belong to more than one fractal box. These parallels call for a systematic investigation, and we have already taken first steps in this direction, for example in \cite{Samsel_2023}. Further research into the interplay between fractality and community structure—especially in networks with hierarchical or overlapping modules—may open up a range of promising and interdisciplinary directions.

\section*{Acknowledgments}


Research was funded by Warsaw University of Technology within the Excellence Initiative: Research University (IDUB) programme (M\L{}, PF).


\end{document}